\documentclass[graybox, envcountchap]{svmult}

\usepackage{mathptmx}        % selects Times Roman as basic font
\usepackage{amsmath}
\usepackage{amssymb}
\usepackage{color}
\usepackage{helvet}          % selects Helvetica as sans-serif font
\usepackage{courier}         % selects Courier as typewriter font
\usepackage{dirtree}
\usepackage{makeidx}        % allows index generation
\usepackage{graphicx}        % standard LaTeX graphics tool
\usepackage{subfig}

\usepackage{multicol}        % used for the two-column index
\usepackage[bottom]{footmisc}% places footnotes at page bottom

\usepackage{hyperref}        %for hyperlinks
\hypersetup{colorlinks=true,urlcolor=blue}

\usepackage{todonotes}

\usepackage[misc]{ifsym}

\makeindex             % used for the subject index
\newcommand{\Neff}{\ensuremath{N_{\rm eff}}}
\newcommand{\DNeff}{\ensuremath{\Delta N_{\rm eff}}}
\newcommand{\Neffstd}{\ensuremath{N_{\rm eff}^{\rm std}}}
\newcommand{\meffs}{\ensuremath{m_{\rm eff,sterile}}}
\newcommand{\Hunits}{km~s$^{-1}$~Mpc$^{-1}$}

\begin{document}

%%%%%%%%%%%%%%%%%%%%%%%%%%%%%%%%%%%%%%%%%%%%%%%%%%%%%%%%%%%%%%%%%

\title{On the dark radiation role in the Hubble constant tension}
\author{Stefano Gariazzo and Olga Mena}
\institute{Stefano Gariazzo (\Letter) \at Istituto Nazionale di Fisica Nucleare (INFN), Sezione di Torino, Via P.\ Giuria 1, I-10125 Turin, Italy, \email{gariazzo@to.infn.it}
\and Olga Mena \at Instituto de F{\'\i}sica Corpuscular  (CSIC-Universitat de Val{\`e}ncia), E-46980 Paterna, Spain, \email{omena@ific.uv.es}}
\maketitle

\abstract{Dark radiation, parameterized in terms of $N_{\rm eff}$, has been considered many times in the literature as a possible remedy in alleviating the Hubble constant ($H_0$) tension. We review here the effect of such an extra dark radiation component in the different cosmological observables, focusing mostly on $H_0$. While a larger value of $N_{\rm eff}$ automatically implies a larger value of the Hubble constant, and one would naively expect that such a simple scenario provides a decent solution, more elaborated models are required. Light sterile neutrinos or neutrino asymmetries are among the \emph{first-order corrections}  to the most economical (\emph{tree-level}) massless dark radiation scenario. However, they are not fully satisfactory in solving the $H_0$ issue. We devote here special attention to \emph{second-order corrections}: some interacting scenarios, such as those with new dark radiation degrees of freedom that exhibit a non-free streaming nature are highly satisfactory alternative cosmologies where to solve the Hubble constant tension.  Models with self-interacting sterile neutrinos and/or majorons, both well-motivated beyond the Standard Model particles, will be discussed along our assessment.}

%%%%%%%%%%%%%%%%%%%%%%%%%%%%%%%%%%%%%%%%%%%%%%%%%%%%%%%%%

\section{Introduction}
\label{sec:1}
Dark radiation in our universe provides a unique window to new mechanisms able to solve many open questions in particle physics and cosmology nowadays. The particle content of the minimal Standard Model (SM) scenario may be enlarged with additional species that feebly couple to ordinary matter, i.e.\ dark radiation.
The putative existence of these species is strongly motivated, both from the pure theoretical and phenomenological/experimental perspectives. The explanation of neutrino masses within the SM and the short-baseline neutrino oscillation anomalies may need the existence of additional sterile neutrino species. The  so-called strong CP problem may imply the existence of axions in our universe, which could contribute significantly to the (observationally required) dark matter component. Last, but not least, Majoron models provide very appealing scenarios where lepton number may be spontaneously broken and rare events such as neutrinoless double beta decay are possible.

Currently, there are several open anomalies that can not be fully understood in the minimal cosmological constant plus cold dark matter ($\Lambda$CDM) scenario. The most significant, $5\sigma$ disagreement is related to the value of the Hubble constant $H_0$ extracted from model-independent, direct measurements of local distances and redshifts in the nearby Universe ($H_0 = 73.04 \pm 1.04$~
\Hunits)~\cite{Riess:2021jrx} and the one indirectly inferred from model-dependent Cosmic Microwave Background (CMB) observations ($H_0 = 67.4 \pm 0.5$~\Hunits)~\cite{Planck:2018vyg}, see also \cite{Riess:2021jrx,Verde:2019ivm,DiValentino:2020zio,DiValentino:2021izs,Schoneberg:2021qvd}. The goal of this manuscript is to analyze the role of dark radiation models in the Hubble constant 
tension. We start in Section~\ref{sec:2} with a brief introduction which summarizes the main properties of the dark radiation component, such as its thermal abundance and its impact on the cosmological observables, focusing mostly on the CMB. Section~\ref{sec:3} describes the most simple models involving dark radiation species ad their role in the $H_0$ conundrum. Section~\ref{sec:4} presents more sophisticated and refined dark radiation scenarios, which may provide a resolution of the $H_0$ issue. We conclude in Sec.~\ref{sec:5}. 

\section{Dark radiation in the early universe}
\label{sec:2}
One of the most important phases of the Universe's evolution is radiation domination.
During this epoch, the largest contribution to the energy density of the Universe comes from relativistic particles,
i.e.~mainly photons, neutrinos and electrons
by the time the temperature of the fluid is below 100~MeV.
During radiation domination, several interesting processes take place:
\textit{i)} neutrinos decouple when the temperature of the plasma is approximately 2~MeV;
\textit{ii)} electrons become non-relativistic shortly after and transfer their energy density to photons while annihilating away,
and \textit{iii)} light nuclei are produced in the Big-Bang Nucleosynthesis (BBN) process until the photon temperature drops below $\sim0.05$~MeV.
Shortly after the end of radiation domination (at $T\sim1$~eV), the last scattering of photons occurs ($T\sim0.3$~eV) and the CMB radiation is produced.
The expansion speed at all these times, proportional to the square root of the radiation energy density,
is crucial because it affects the observables we measure today.

After electron-positron pairs annihilate into photons,
neutrinos and photons are the only SM particles behaving as radiation.
Other relativistic particles may exist, as we will discuss in the following.
Such non-standard particles are normally grouped under the name "dark radiation", since they do not take part into electroweak interactions.
The amount of the dark radiation energy density, $\rho_{\rm R}$,
can be conveniently parameterized by means of the
effective number of relativistic neutrino-like species, \Neff, by
\begin{equation}
\rho_{\rm R}
=
\rho_\gamma
\left(
1+
\frac{7}{8}
\left(\frac{4}{11}\right)^{4/3}
\Neff
\right)\,,
\end{equation}
where $\rho_\gamma$ represents the photon energy density.
If only three standard neutrinos which underwent instantaneous decoupling exist, \Neff\ would be equal to 3.
In case the neutrino energy density is different from the instantaneous decoupling one,
\Neff\ may deviate from 3 even in presence of only three standard neutrino families.
The presence of additional contributions to the radiation energy density, moreover,
would correspond to an increased value of \Neff.
In other words, $\Neff \neq 3$ might originate either because of new degrees of freedom which have nothing to do with standard neutrinos,
or from a non-standard momentum distribution for the three neutrinos.

Our theoretical knowledge of their decoupling,
however, states that standard neutrinos did not undergo instantaneous decoupling
and that their momentum distribution function slightly deviates from a pure Fermi-Dirac one.
When taking into account the momentum dependence of neutrino-electron interactions and
the evolution of the thermal plasma around electron-positron annihilation,
the standard neutrino energy density is computed to be $\Neffstd=3.044$ \cite{Akita:2020szl,Froustey:2020mcq,Bennett:2020zkv}, see also~\cite{Cielo:2023bqp}.
This number was previously claimed to be a bit higher \cite{Mangano:2005cc,deSalas:2016ztq},
but state-of-the-art calculations confirm that the (theoretical and numerical) error on this value is at the level of $10^{-4}$.
The presence of new physics related to the three standard neutrinos,
such as non-standard interactions (NSI, see e.g.~\cite{Farzan:2017xzy}) between neutrinos and electrons \cite{Du:2021idh,deSalas:2021aeh},
or a non-unitary neutrino mixing matrix \cite{Gariazzo:2022evs}
can lead to deviations from \Neffstd.
These effects are rather small and do not significantly impact the $H_0$ value,
although they may be tested by next-generation CMB measurements \cite{Ade:2018sbj,CMB-S4:2016ple}.
As we shall see, the situation changes significantly if we consider the presence of additional particles
(axions, sterile neutrinos, and so on)
or more complicated neutrino interactions, for example with dark matter.
In such cases, it is common to have much larger contributions to \Neff, usually denoted as
$\DNeff\equiv\Neff-\Neffstd$. An exhaustive description of the effect of \DNeff\ on the different cosmological observables can be found in Ref.~\cite{Archidiacono:2013fha}. Current limits on $\Delta N_{\rm eff}$ arise primarily from observables at two epochs: \textit{(i)} at the BBN period, and, \textit{(ii)} at the Recombination epoch. Here we briefly summarize the main impact of this parameter on both BBN light element abundances and on the CMB power spectrum, as the impact on the matter power spectrum is subdominant.

BBN refers to the formation of the first light nuclei (heavier than the lightest isotope of hydrogen) in the very first three minutes of our universe's lifetime. BBN abundances are computed by employing a solid understanding of the nuclear interactions involved in the production of elements,  providing a natural laboratory where to test extensions of the SM of elementary particles that involve additional relativistic species \DNeff.  Any extra contribution to the dark radiation of our universe will increase the expansion rate $H(z)$ and will shift towards higher temperatures the freeze-out epoch of the weak interactions, implying a higher neutron-to-proton ratio and therefore a larger fraction of primordial Helium and Deuterium (as well as a higher fraction of other primordial elements) with respect to hydrogen. This makes BBN a powerful tool for constraining the total amount of relativistic species and other beyond-the-SM physics frameworks: given a concrete model, we can solve numerically the set of differential equations that regulate the nuclear interactions in the primordial plasma, see e.g.~\cite{Pisanti:2007hk,Consiglio:2017pot,Gariazzo:2021iiu}, compute the light element abundances and compare the results to the values inferred by astrophysical and cosmological observations. Given current uncertainties, the standard BBN predictions show a good agreement with direct measurements of primordial abundances~\cite{Pitrou:2020etk,Mossa:2020gjc,Pisanti:2020efz,Yeh:2020mgl} limiting $\DNeff\lesssim 0.3-0.4$ at 95\% CL. Notice also that the BBN predictions for the Helium abundance can impact the CMB angular spectra because they can be used to estimate the baryon energy density through a simple formula~\cite{Serpico:2004gx}:
\begin{equation}
    \Omega_b h^2 = \frac{1 - 0.007125\ Y_p^{\rm BBN}}{273.279}\left(\frac{T_{\rm CMB}}{2.7255\ \mathrm{K}}\right)^3 \eta_{10} \ .
    \label{eq.Yp}
\end{equation}
Here, $\eta_{10} \equiv 10^{10}n_b/n_\gamma $ is the photon-baryon ratio today, $T_{\rm CMB} $ is the CMB temperature at the present time and $Y_p^{\rm BBN} \equiv 4 n_{\rm He}/n_{b}$ is the Helium \textit{nucleon fraction}, defined as the ratio of the 4-Helium number density to the total baryon one. 

Concerning the CMB temperature power spectrum, first of all, varying \Neff\ changes the redshift of the matter radiation equivalence, $z_{\rm eq}$, inducing an enhancement of the early Integrated Sachs Wolfe (ISW) effect which increases the CMB spectrum around the first acoustic peak. Nevertheless, this is a sub-dominant effect.
The authors of Ref.~\cite{Hou:2011ec} carefully explained that the most relevant impact of changing \Neff\ is located at high multipoles, i.e.\ at the damping tail. 
If \DNeff\ increases, the Hubble parameter $H$ during radiation domination will increase as well. The delay in matter radiation equality will also modify the sound speed and the comoving sound horizon, proportional to $1/H$, which reads as:
\begin{displaymath}
r_{\rm s} = \int_0^{\tau'} d\tau c_{\rm s} (\tau) = \int_0^a \frac{da}{a^2 H} c_{\rm s}(a)~,
\end{displaymath}
and it is proportional to the inverse of the expansion rate $r_{\rm s}\propto1/H$. The consequence is a reduction in the angular scale of the acoustic peaks $\theta_{\rm s}=r_{\rm s}/D_{\rm A}$, where $D_{\rm A}$ is the angular diameter distance, causing a horizontal shift of the peak positions towards higher multipoles. Furthermore, a vertical shift also affects the amplitude of the peaks at high multipoles, where the ISW effect is negligible, and it is related to the so-called Silk damping.
This results from the fact that the baryon-photon decoupling is not an instantaneous process, leading to a diffusion damping of oscillations in the plasma.
If decoupling starts at $\tau_{\rm d}$ and ends at $\tau_{\rm ls}$, during $\Delta\tau$
the radiation free streams on scale $\lambda_{\rm d}=\left(\lambda\Delta\tau\right)^{1/2}$
where $\lambda$ is the photon mean free path
and $\lambda_{\rm d}$ is shorter than the thickness of the last scattering surface.
As a consequence, temperature fluctuations on scales smaller than $\lambda_{d}$ are damped,
because on such scales photons can spread freely both from overdensities and underdensities. The overall result is that the damping angular scale $\theta_{\rm d}=r_{\rm d}/D_{\rm A}$ is proportional to the square root of the expansion rate $\theta_{\rm d}\propto\sqrt{H}$ and consequently it increases with \DNeff\, which therefore induces a suppression of the peaks and a smearing of the oscillations that intensifies at the CMB damping tail.

The three aforementioned effects caused by a non-zero \DNeff\ (redshift of equivalence, the size of the sound horizon at recombination, and the damping tail suppression) can be easily compensated by varying other cosmological parameters, including the Hubble constant $H_0$. Notice that the horizontal shift towards smaller angular scales caused by an increased value of \Neff\ can be compensated by decreasing $D_{\rm A}$, which can be automatically satisfied by increasing $H_0$. The effect of \Neff\ on the damping tail is however more difficult to mimic via $H_0$, as it is mostly degenerated with the helium fraction which enters directly in $r_{\rm d}$,  i.e.\ the mean square diffusion distance at recombination via $n_e$, the number density of free electrons. 

There is however one effect induced by \Neff\ which cannot be mimicked by other cosmological parameters: the neutrino anisotropic stress~\cite{Bashinsky:2003tk,Hannestad:2004qu}, related to the fact that neutrinos are free-streaming particles propagating at the speed of light, faster than the sound speed in the photon fluid, suppressing the oscillation amplitude of CMB modes that entered the horizon in the radiation epoch. The effect on the CMB power spectrum is therefore located at scales that cross the horizon before the matter-radiation equivalence by an increase in power of $5/(1+\frac{4}{15}f_\nu)$ \cite{Hu:1995en}, where $f_\nu$ is the fraction of radiation density contributed by free-streaming particles.

All in all, our current knowledge confirms that \Neff\ is close to 3 as measured
by CMB observations ($\Neff=2.99^{+0.34}_{-0.33}$ at 95\% confidence level (CL) \cite{Planck:2018vyg})
or BBN abundances (e.g.~$\Neff=2.87^{+0.24}_{-0.21}$ at 68\% CL \cite{Consiglio:2017pot}) independently.

\section{\emph{Tree-level} dark radiation solutions to the $H_0$ tension}
\label{sec:3}
Let us now analyse the simplest dark radiation candidate that can be used to extend the standard three-neutrino picture and discuss their effect on the $H_0$ tension.

\subsection{Sterile neutrinos}
\label{subsec:3.1}
One simple dark radiation candidate is the sterile neutrino,
intended as a right-handed fermion which cannot take part into SM interactions
but that participates in neutrino oscillations.
In order to be considered as dark radiation, the sterile neutrino must be relativistic in the early universe, indicatively at the time of BBN or CMB epochs.
This excludes sterile neutrinos with masses much larger than the MeV scale, which are usually proposed in order to build a seesaw model and explain the smallness of active neutrino masses, see e.g.~\cite{King:2015sfk,Miranda:2016ptb,Xing:2020ald}.
Sterile neutrinos at the keV scale can play a significant role in the universe as warm dark matter~\cite{Boyarsky:2018tvu,Adhikari:2016bei}, but since they become non-relativistic during BBN, they are not considered as dark radiation components.
When considering masses at the eV scale, sterile neutrinos would become non-relativistic approximately at the time of matter-radiation equality, thus being relativistic during the entire radiation-dominated era,
although already non-relativistic at CMB decoupling ($T\sim0.3$~eV).
Sterile neutrinos corresponding to much smaller mass splittings with respect to active neutrinos (mass splittings smaller than the solar or atmospheric ones) are also considered in quasi-Dirac or pseudo-Dirac scenarios, see e.g.~\cite{Anamiati:2019maf, Carloni:2022cqz}.
In this case, they may remain relativistic until today, depending on the mass of the lightest neutrino.
In the following, however, we will focus primarily on the eV-scale sterile neutrino.

Sterile neutrinos at the eV scale have been introduced to give a possible neutrino-oscillation explanation to the so-called Short-Baseline (SBL) anomalies~\cite{Gariazzo:2015rra,Giunti:2019aiy,Boser:2019rta,Archidiacono:2022ich},
which originally included the excess appearance of electron antineutrinos measured by LSND~\cite{LSND:2001aii}, later confirmed also by MiniBooNE~\cite{MiniBooNE:2008yuf,MiniBooNE:2018esg},
and the anomalous disappearance of electron neutrinos at Gallium experiments~\cite{Abdurashitov:2005tb,Giunti:2010zu}
and of electron antineutrinos from nuclear reactors~\cite{Mention:2011rk}.
The original anomalies have been studied at different experimental probes and at the theoretical level for more than ten years.
Here we will briefly summarize the present status.

Concerning the appearance channel, the LSND results have been tested by several experiments apart from MiniBooNE.
None of these experiments has observed an anomalous signal, including
KARMEN~\cite{Armbruster:2002mp}
and OPERA~\cite{OPERA:2018ksq},
which therefore put only upper bounds on active-sterile mixing angles.
In more recent years, the MicroBooNE experiment is testing the MiniBooNE anomaly,
and the first results seem to indicate that the excess of events is not compatible with
active-sterile neutrino oscillations~\cite{MicroBooNE:2021zai,MicroBooNE:2022wdf}.

In the disappearance channel,
we can separate the electron and muon (anti)neutrino channels.
In the former case,
the Gallium and reactor anomalies have been studied both experimentally and theoretically.
Recent revisitations of the Gallium cross sections appear to reduce the significance of the Gallium anomaly with respect to the original calculations~\cite{Giunti:2022btk,Giunti:2022xat},
but the BEST experiment observed a disappearance with stronger significance~\cite{Barinov:2021asz,Barinov:2022wfh}
with respect to previous GALLEX and SAGE results,
see also \cite{Elliott:2023cvh}.
If neutrino oscillations are responsible for the BEST results,
part of the preferred region is in tension with SBL reactor antineutrino experiments~\cite{STEREO:2019ztb,PROSPECT:2020sxr,DANSS:2018fnn,RENO:2020hva}.
On the other hand,
although most of the reactor experiments do not confirm the original reactor anomaly~\cite{STEREO:2019ztb,PROSPECT:2020sxr,DANSS:2018fnn,Alekseev:2022tcz,NEOS:2016wee,RENO:2020hva},
the Neutrino-4 experiment claims a best-fit at a high mass splitting and mixing angle \cite{Serebrov:2023onj},
whose significance may be overestimated due to the treatment of the $\chi^2$~\cite{Giunti:2021iti}.
Neutrino-4 oscillation results are perfectly compatible with BEST constraints,
which however have a much broader allowed range for the new mass splitting.
Recent theoretical calculations of the reactor antineutrino flux~\cite{Estienne:2019ujo,Kopeikin:2021rnb,Perisse:2023efm}
differ slightly from the two that produced the original reactor anomaly~\cite{Mueller:2011nm,Huber:2011wv},
which is therefore significantly reduced or even disappeared if the most updated flux estimations are considered~\cite{Giunti:2021kab,Perisse:2023efm}.

Finally, concerning muon (anti)neutrino disappearance,
most of the experiments, for example MINOS/MINOS+~\cite{MINOS:2017cae}, only put upper limits on the mixing parameters because no SBL oscillations are observed.
Only in recent years, IceCube reported a mild preference in favor of active-sterile neutrino oscillations by analysing 8 years of muon neutrino events~\cite{IceCube:2020phf},
with a broad uncertainty on the preferred mass splitting and mixing angles.

Assuming that all the above-mentioned anomalies are generated by active-sterile neutrino mixing,
when one attempts to combine all these appearance and disappearance results
in order to produce a global fit of active-sterile oscillation parameters,
it seems impossible to reconcile the strong upper bounds from disappearance experiments
with the observations at appearance probes.
The tension was already strong a few years ago~\cite{Gariazzo:2017fdh,Dentler:2018sju,Giunti:2019aiy},
at the point that a combination of appearance and disappearance constraints has no statistical meaning.
In summary, the sterile neutrino searches from different probes seem unable to pin out a single preferred region for the active-sterile oscillation parameters,
and possibly a consistent explanation of the anomalies requires different kinds of new physics.
In the following, therefore,
we will consider the potential impact of eV-scale sterile neutrinos on the cosmological environment and ignore the problems in terrestrial searches of sterile neutrinos.

The thermalization of sterile neutrinos in the early universe has been studied for a long time, see e.g.~\cite{Barbieri:1989ti,Barbieri:1990vx,Kainulainen:1990ds,Enqvist:1990ek,Enqvist:1991qj,Dolgov:2003sg,Cirelli:2004cz}, or \cite{Dolgov:2002wy} for a review of early studies.
The most important result in these early calculations is to show that the production of sterile neutrinos occurs mostly in a non-resonant way through neutrino oscillations.
Reference~\cite{Dolgov:2003sg} provides a rather simple formula that determines the relation between active-sterile mixing parameters and the approximated value of \DNeff\ they generate.
In more recent years, precise numerical calculations have been developed in order to determine the contribution of eV-scale sterile neutrinos
to \DNeff.
Calculations have been first implemented in a
"one active plus one sterile neutrino" scenario but keeping the full momentum dependence, see e.g.~\cite{Hannestad:2012ky,Hannestad:2013wwj,Hannestad:2015tea},
and then extended to include multiple flavors (two active and one sterile neutrino)~\cite{Mirizzi:2012we,Mirizzi:2013gnd,Saviano:2013ktj}.
Finally, more accurate calculations at the numerical level have been proposed in recent years~\cite{Gariazzo:2019gyi,Mastrototaro:2021wzl}.

The amount of sterile neutrinos produced through oscillations in the early universe depends significantly on the values of the active-sterile mixing parameters.
Because the dense plasma blocks neutrino oscillations at very early times by maintaining
neutrinos into a flavor state, oscillations cannot take place until rather late.
Since this effect depends on the oscillation frequency, higher mass splittings correspond to an earlier start of oscillations,
that in turn means that there is more time to produce sterile neutrinos.
Lower mass splittings, instead, generate oscillations that are blocked for a longer time,
and are less efficient in producing sterile states.
On the other hand, a large mixing angle allows to have a faster conversion between active and sterile flavors.
Finally, it has been noticed that the new active-sterile mixing angles are equivalent at the time of producing neutrino oscillations in the early universe.
More specifically, for a $\Delta m_{41}^2\sim1$~eV, it is sufficient to have any of the mixing angles larger than $\sim10^{-3}$ to generate a sterile state in full thermal equilibrium with the active neutrinos, or $\DNeff\simeq1$ \cite{Gariazzo:2019gyi}.

This is of course in tension with the constraints we obtain from CMB and BBN observables.
Cosmological analyses normally parameterize the presence of sterile neutrinos by means of its contribution
\DNeff\ to the effective number of neutrinos when relativistic,
and its contribution to the matter energy density, proportional to the effective mass $\meffs$, when it becomes non-relativistic.
In case of non-resonant production of the sterile state \cite{Dodelson:1993je}, the relation 
$\meffs=m_s\DNeff$ applies, where $m_s$ is the sterile neutrino physical mass.
From the sterile neutrino search,
Planck obtains upper limits $\Neff<3.30$ and $\meffs<0.652$~eV at 95\% CL (TTTEEE+lensing+BAO) \cite{Planck:2018vyg}.
Notice that in this case, the analysis also yields $H_0=67.8^{+1.3}_{-1.2}$~\Hunits\ at 95\% CL,
meaning that the Hubble tension is not alleviated.
The constraints are particularly strong thanks to the combined effect of \Neff\ and $H_0$ on both temperature and polarization.
If one only considers the TT+lensing+BAO dataset, Planck reports
$\Neff<3.52$, $\meffs<0.551$~eV and $H_0=68.2^{+2.1}_{-1.7}$~\Hunits,
which is still in tension with a fully thermalized fourth neutrino, but with a slightly reduced significance.
Notice that the bounds on $\Neff$ being smaller than 4 as reported by Planck are also confirmed by ACT and SPT, which respectively yield
$\Neff<3.38$, $\meffs<0.561$~eV
and $\Neff<3.86$, $\meffs<0.232$~eV
when analysed in combination with WMAP 9-year data and other low-redshift probes \cite{DiValentino:2023fei}.
Notice that in the SPT case, the bound on \meffs\ is much stronger, while the bound on \Neff\ is relaxed.
This arises from the fact that heavier states are allowed only if their contribution to \Neff\ is rather small (otherwise their contribution to the non-relativistic energy density, proportional to $\meffs$, would be too high),
while for lighter states a larger value of \Neff\ is allowed because these particles mostly act as radiation.
Even in the SPT case, however, these results show that none of the available cosmological data sets currently allows for the presence of a fully thermalized sterile neutrino,
and the allowed parameter ranges do not permit a full solution of the $H_0$ tension.

\subsection{Neutrino asymmetries}
\label{subsec:3.2}
In the standard big-bang model, the neutrino asymmetry $\eta_{\nu_\alpha}\equiv (n_{\nu_\alpha}- n_{\bar{\nu}_\alpha})/n_\gamma$ is assumed to be negligibly small, comparable to the tiny-but-crucial baryon asymmetry. 
That is, while standard baryogenesis models involving sphalerons suggest that $\eta_{\nu_\alpha}$ should be of the order
of the baryon asymmetry $\eta_B = n_B/n_\gamma \simeq 6.1 \times 10^{-10}$~\cite{Fields:2019pfx,ParticleDataGroup:2022pth}, other proposed models~\cite{Gu:2010dg,March-Russell:1999hpw,McDonald:1999in} manage to combine a large lepton asymmetry with the value of $\eta_B$.   The initial asymmetry in a given neutrino flavour $\alpha$, defined as the difference between the neutrino and
antineutrino comoving densities, is related to the neutrino chemical potential $\mu_\alpha$. It is possible to compute the constraints on this parameter, or, as it is usually done in the literature, on the dimensionless degeneracy parameter $\xi_\alpha \equiv \mu_\alpha/T_\nu$.  
A non-vanishing value of the electron neutrino chemical potential $\xi_e$ affects the neutron/proton freeze-out, modifying BBN predictions~\cite{Froustey:2019owm}.  Notice that a priori, the individual neutrino flavour chemical potentials can have different values.
Depending on the neutrino mixing
angles as well as on the initial values of the neutrino chemical potentials there could be or not an equilibration of the different $\xi_\alpha$~\cite{Barenboim:2016lxv}~\footnote{%
When assuming a unique common degeneracy parameter $\xi$ for the three neutrino degeneracy parameters due to neutrino oscillations~\cite{Mangano:2011ip,Dolgov:2002ab,Wong:2002fa}, one can derive bounds from BBN observations~\cite{Simha:2008mt,Fields:2019pfx}, from
CMB data~\cite{Planck:2018vyg,Oldengott:2017tzj}, or from a combination of the former two measurements, providing $\xi= 0.001 \pm 0.016$~\cite{Pitrou:2018cgg,Froustey:2021azz}.}. A significant lepton asymmetry in the early universe, represented by a nonvanishing chemical potential $\xi_{\alpha_i}$, for thermal distributions of two light mass eigenstates $\nu_i$, will contribute to the dark radiation of the universe as~\cite{Barenboim:2016lxv}
\begin{equation}
\DNeff = \sum_{i=1,2} \frac{15}{7}\left(\frac{\xi}{\pi}\right)^2 \left(2+\left(\frac{xi}{\pi}\right)^2\right)~.
\end{equation}
The authors of Ref.~\cite{Barenboim:2016lxv} show that CMB data alone lead to a 95\%~CL limit of $\xi<0.77$, associated to $\DNeff=0.3$ and $H_0=67.71\pm 0.95$~\Hunits, and therefore not alleviating the $H_0$ tension. However, a combination with Supernovae I data provides weak evidence for a non negligible value of $\xi$, together with a $95\%$~CL lower bound for the Hubble parameter of $H_0>69.8$~\Hunits, larger than the value without Supernovae observations and in a better agreement with local measurements of the Hubble constant.

\subsection{Axions}
\label{subsec:3.3}

Axions are the dynamical pseudo-scalar fields providing one of the most compelling solutions to the strong CP problem~\cite{Peccei:1977hh,Peccei:1977ur}. In addition, axions also may be considered as a natural candidate for the dark matter component in our universe. Apart from non-thermal mechanisms, axions can be copiously produced in the early Universe also via scattering and decays of particles belonging to the primordial bath~\cite{Turner:1986tb,Berezhiani:1992rk,Brust:2013ova,Baumann:2016wac,DEramo:2018vss,Arias-Aragon:2020qtn,Arias-Aragon:2020shv,Green:2021hjh,DEramo:2021usm},   contributing to the radiation energy density \Neff\ in the same way as massive neutrinos do when they are still relativistic.
Eventually, axions become non-relativistic and provide a hot and sub-dominant dark matter component. The presence of such a cosmic axion background can leave distinct and detectable imprints, and current cosmological data put bounds on the axion mass and interactions~\cite{Hannestad:2005df,Melchiorri:2007cd,Hannestad:2007dd,Hannestad:2008js,Hannestad:2010yi,Archidiacono:2013cha,Giusarma:2014zza,DiValentino:2015zta,DiValentino:2015wba,Archidiacono:2015mda,Giare:2020vzo,DEramo:2022nvb,Giare:2021cqr,DiValentino:2022edq}. Axion couplings are proportional to the mass of the axion itself, and thermal production channels are efficient only if the axion mass is large enough. On the contrary, the cold axion dark matter density is a decreasing function of the mass. Therefore, a significant thermal  population of axions would be possible only if cold axions provide a sub-dominant component to the observed cosmic dark matter abundance. In terms of  $f_a$, known as the axion decay constant,  the axion mass reads as~\cite{Bardeen:1978nq,GrillidiCortona:2015jxo,Gorghetto:2018ocs}
\begin{equation}
m_a \simeq 5.7 \, \mu{\rm eV} \, \left( \frac{10^{12} \, {\rm GeV}}{f_a} \right) \ .
\label{eq:ma}
\end{equation}

The landscape of axion models is broad~\cite{Kim:2008hd,DiLuzio:2020wdo}. Nevertheless, it is possible to divide them into two main classes according to the origin of the color anomaly: the Kim-Shifman-Vainshtein-Zakharov (KSVZ)~\cite{Kim:1979if,Shifman:1979if} and Dine-Fischler-Srednicki-Zhitnitsky (DFSZ)~\cite{Dine:1981rt,Zhitnitsky:1980tq} frameworks, leading to the very same axion mass, see Eq.~(\ref{eq:ma}). However, the interactions with electroweak gauge bosons and SM fermions  are different. Scatterings are the only thermal production channel available for the minimal KSVZ and DFSZ frameworks due to the flavor-conserving axion interactions. If flavor-violating couplings are allowed then decays also contribute~\cite{DEramo:2021usm}. Initially, axions may belong to the thermal bath if during or after inflationary reheating there are efficient mechanisms to produce them. Even if they are not present at the onset of the radiation domination epoch, they are produced by thermal scatterings and their interaction strength could bring them to equilibrium. 
A quick method to estimate the axion relic density is via the \textit{instantaneous decoupling approximation}, assuming that axions at some point reach thermal equilibrium, and that they suddenly decouple when the bath had a temperature $T_D$ identified by the relation 
\begin{equation}
\Gamma_a(T_D) = H(T_D) \ .
\label{eq:TD}
\end{equation}
The asymptotic axion comoving density results in
\begin{equation}
Y_a^\infty = Y_a(T \lesssim T_D) = \frac{45 \zeta(3)}{2 \pi^4 g_{\star s}(T_D)}~, 
\label{eq:Yainf}
\end{equation}
with $\zeta(3) \simeq 1.2$ the Riemann zeta function. The asymptotic  comoving density depends on the decoupling temperature only through the factor $g_{\star s}(T_D)$ in the denominator. Axions decoupling at later times are a more significant fraction of the thermal bath and therefore their comoving density is larger.  Notice that the criterion in Eq.~(\ref{eq:TD}) leads to a convenient estimate of the decoupling epoch but it is far from being rigorous. As it is manifest from Eq.~(\ref{eq:Yainf}), the final axion abundance is sensitive to the detailed value of the decoupling temperature only if $g_{\star s}$ is changing around the decoupling time, see Ref.~\cite{DEramo:2022nvb}, which provides  a robust computation of the axion relic density calculation in terms of the Boltzmann equation, in order to find solid cosmological bounds on the axion mass, finding that the actual axion relic density is substantially larger than previously estimated. Once the relic density is computed, it is possible to evaluate the axion contribution to the effective number of additional neutrino species via the relation $\Delta N_{\rm eff}\simeq 75.6\,\left(Y_a^{\infty}\right)^{4/3}$, assuming that the phase-space distribution is thermal. Therefore,  as long as thermal axions remain relativistic particles ($T_a\gg m_a$), they behave as radiation in the early Universe and their cosmological effects are those produced via their contribution to the effective number of neutrino species \Neff. By means of BBN
light element abundances data, Ref.~\cite{DEramo:2022nvb}  finds for the KSVZ axion $\DNeff<0.33$ and an axion mass bound
$m_a < 0.53$~eV (i.e., a bound on the axion decay constant $f_a > 1.07 \times 10^7$~GeV), both at 95\%~CL.
These BBN bounds are improved to $\DNeff<0.14$ and $m_a < 0.16$~eV ($f_a > 3.56 \times 10^7$~ GeV) if a
prior on the baryon energy density from CMB data is assumed.
When cosmological observations from the CMB temperature, polarization and
lensing from the Planck satellite  are combined with large scale structure data  limits of  $\DNeff < 0.23$ and 
$m_a < 0.28$~eV are found, both at 95\%~CL. 
Very similar results are quoted for the DFSZ axion. The authors of Ref.~\cite{DEramo:2018vss} showed that a model in which hot axions are produced from the coupling with muons leads to an alleviation of the Hubble tension at $3\sigma$.

\section{\emph{Higher order} dark radiation corrections to the $H_0$ tension}
In the following, we shall present proposed scenarios  which imply extra interactions beyond the SM paradigm, involving also in some cases exotic particles, such as sterile neutrino species, the dark matter component, or Majorons. 
\label{sec:4}
\subsection{Interacting scenarios}
\label{subsec:4.1}
Both neutrinos and dark matter provide evidence for physics beyond the SM of elementary particles and possible interactions among them can induce important changes in the canonical evolution within the minimal $\Lambda$CDM model, see Refs.~\cite{Mangano:2006mp,Serra:2009uu,Diamanti:2012tg,Wilkinson:2014ksa,Olivares-DelCampo:2017feq,Stadler:2019dii,Mosbech:2020ahp,
Escudero:2018mvt,Escudero:2020dfa}. In particular, in  these models, the value of $\Neff$ \ will generically be increased at photon decoupling while will remain unchanged during the BBN period. Since the effective number of relativistic degrees of freedom is proportional to the ratio of the neutrino and photon temperatures, if the photon temperature $T_\gamma$ is reduced, or, alternatively, the neutrino temperature is increased, $\Neff$ will be larger than in the canonical scenario. If there exists a dark matter particle that remains in thermal equilibrium  with neutrinos after their weak decoupling processes, these interactions will effectively reheat the neutrino sector with respect to the electromagnetic plasma. This neutrino reheating will take place after standard neutrino decoupling once the dark matter particles are no longer relativistic, implying they should have a mass of tens of MeV~\cite{Boehm:2012gr,Boehm:2013jpa}. It has been explored if such a scenario could alleviate the $H_0$ tension~\cite{DiValentino:2017oaw,Mosbech:2020ahp}.
Including Planck polarization data, however, dilutes such a possibility, although this very same scenario is able to significantly relax the lower bounds on the value of the clustering parameter $\sigma_8$ inferred in the context of $\Lambda$CDM from the Planck data, leading to
agreement within $1-2\sigma$ with weak lensing estimates of the $\sigma_8$ parameter.

\subsection{Self-interacting active neutrinos}
\label{subsec:4.2}
Free-streaming neutrinos travel supersonically through the photon-baryon  plasma at early times, inducing a net phase shift in the CMB power spectra towards larger scales (smaller multipoles), leading to a physical size of the photon sound horizon at last scattering that is slightly larger~\cite{Bashinsky:2003tk,Follin:2015hya,Baumann:2015rya,Baumann:2017lmt,Choi:2018gho}. 
Self-interacting neutrinos, see e.g.~\cite{Berryman:2022hds,Taule:2022jrz} shift the power spectrum towards smaller scales and boost their fluctuation amplitude, reducing the physical size of photon sound horizon at last scattering: a smaller value of the angular diameter distance would be required, implying a higher value of $H_0$~\cite{Kreisch:2019yzn}.
If the mass of the mediator is heavy, the process can be expressed in terms of a four-fermion interaction $G_{\rm eff} \nu \bar{\nu}\nu \bar{\nu} $~\cite{Kreisch:2019yzn}, with $G_{\rm eff} \gg G_{\rm F}$, the former governing weak-interaction processes.
Neutrinos experience self-scatterings after weak decoupling, and increasing $G_{\rm eff}$ delays neutrino free-streaming. The authors of Ref.~\cite{Kreisch:2019yzn} found a strongly interacting mode with $G_{\rm eff} = 2.5 ^{+0.8}_{-0.5} \times 10^4 \ \rm{GeV}^{-2}$ and associated it to a much larger value of $H_0$ than in the standard $\Lambda$CDM model, $H_0 = 72.3 \pm 1.4$~\Hunits.
This very same 
strongly interacting neutrino cosmology prefers $\Neff = 4.02 \pm 0.29$.
Despite being a very interesting result, the former analysis was restricted to Planck observations of temperature power spectrum data:  the authors of Ref.~\cite{RoyChoudhury:2020dmd} realized that when high polarisation data from the Planck 2018 release are included in the fit, the strongly interacting neutrino mode was disfavoured, even if it could not be completely excluded.
As a consequence, strong
neutrino self-interactions do not lead to a high value of the Hubble constant and therefore such a  model is
not a viable solution to the current $H_0$ discrepancy when considering the full Planck 2018 data, see also Refs.~\cite{Brinckmann:2020bcn,Das:2021guu,Brinckmann:2022ajr,Sandner:2023ptm,Escudero:2022rbq,Taule:2022jrz,Kreisch:2022zxp,Venzor:2023aka}.

Among the classes of theoretical neutrino model building, the self-interacting scenario discussed here
includes an important class of models related to the explanation of the smallness of neutrino masses. Why are the neutrino masses within the SM much smaller than those of the charged fermions in the very same family?
One of the most elegant and complete benchmarks is the so-called seesaw mechanism~\cite{King:2015sfk,Miranda:2016ptb,Xing:2020ald}, in which additional right handed neutrinos are added via a Majorana mass term, naturally suppressing the (light, active) neutrino mass scale. These models imply a lepton number spontaneous symmetry breaking, with an associated pseudo-Goldstone boson, the \emph{Majoron}~\cite{Chikashige:1980qk,Chikashige:1980ui}, a light scalar with very weak interactions with visible matter~\cite{Barger:1981vd,Akita:2023qiz}.
Nevertheless, Majoron-neutrino interactions will change the cosmological observables at both the BBN and CMB periods, as they contribute to both \DNeff\ and to the neutrino anistotropic stress, reducing neutrino free-streaming~\cite{Bashinsky:2003tk,Chacko:2003dt,Lattanzi:2007ux,Lattanzi:2013uza,Biggio:2023gtm}.
With Planck 2015 data, the Majoron models were shown to predict a higher value of the $H_0$ parameter, especially when $\Neff$ is free in the analysis \cite{Forastieri:2015paa}.
Updated studies demonstrated that the flexibility in explaining the $H_0$ tension was
reduced when considering non-CMB data together with Planck \cite{Forastieri:2019cuf},
as the $H_0$ values obtained in the combined analysis of Planck 2015 TTTEEE plus external data is $68.13 \pm 0.48$~\Hunits.
Considering Planck 2018 data and an updated local measurement of $H_0$, instead,
the authors of Ref.~\cite{Escudero:2019gvw} show that Majorons could alleviate the Hubble constant tension: as these particles contribute naturally to the dark radiation component of the universe, with $\DNeff \sim 0.11$, they are able to reduce by a significant level the $H_0$ discrepancy.
The combined fit of CMB, BAO and $H_0$ measurements improves by $\Delta\chi^2=-12.2$~\cite{Escudero:2019gvw} when considering the self-interacting scenario with respect to the standard $\Lambda$CDM model, see also Refs.~\cite{Arias-Aragon:2020qip,Huang:2021dba,Escudero:2021rfi}, although the CMB $\chi^2$ alone is slightly worsened. 
Exploiting the very same mechanism, i.e.\ a very light scalar Majoron coupled to neutrinos, the authors of Ref.~\cite{Barenboim:2020dmg} found a better agreement among high redshift estimates and local,  direct measurements of $H_0$.

\subsection{Self-interacting sterile neutrinos}
\label{subsec:4.3}
Another class of self-interacting scenarios involves the coupling between a sterile neutrino and a new pseudo-scalar particle, see e.g.~\cite{Archidiacono:2022ich} and references therein.
This secret interaction would induce a large matter potential that suppresses active-sterile
oscillations in the early universe.
During the universe expansion, when the matter potential becomes similar to the vacuum oscillation frequency,
the sterile neutrino may encounter a resonance \cite{Forastieri:2017oma}.
Depending on the mass of the mediator, the resonance can occur at different times.
If the new particle is much lighter than the sterile neutrino, the resonance occurs much later than BBN, leaving the abundances unaffected, and the sterile state is produced through vacuum neutrino oscillations.
At some point of the evolution, moreover, the sterile neutrino can decay into the new mediator
and the mass bounds are avoided
\cite{Archidiacono:2014nda,Archidiacono:2015oma,Archidiacono:2016kkh}.
If the mediator is heavier than the MeV scale
\cite{Saviano:2014esa,Mirizzi:2014ama},
instead, the sterile neutrino might be produced through a resonant process.
If this happens before BBN, the light element abundances would be significantly altered.
If the sterile neutrino production occurs after BBN, even ignoring the resonant mechanism, a copious abundance of sterile neutrinos is generated by active-sterile conversions.
In such case, the scenario would be disfavored by cosmological bounds on the neutrino energy density.

When considering a light pseudoscalar, the most recent analyses performed with Planck data \cite{Archidiacono:2020yey}
show that the self-interacting model naturally corresponds to a high value for $H_0$,
namely $H_0=71.6^{+1.1}_{-1.6}$~\Hunits,
which is very close to the local observed value.
This result corresponds to
a slightly better fit ($\Delta\chi^2=-1.0$) of PlanckTTTEEE + $H_0$ constraints than a simple sterile neutrino scenario,
although the CMB $\chi^2$ alone is slightly worsened with respect to the non-interacting case.
When fitting PlanckTTTEEE+lensing+BAO data, the model allows to obtain $H_0=70.0^{+0.7}_{-1.1}$~\Hunits,
which is still much higher than the CMB prediction in the $\Lambda$CDM model.
When considering high-multipoles CMB data, the light pseudoscalar scenario is instead preferred
over the $\Lambda$CDM model by ACT data alone ($\Delta\chi^2=-5.3$),
moreover with a preferred value  of $H_0=72.7^{+2.0}_{-2.8}$~\Hunits,
although the combination of Planck+ACT prefers the simpler $\Lambda$CDM model
with a $\Delta\chi^2=17.4$ against the pseudoscalar case
and a lower value of $H_0=70.6^{+0.7}_{-0.9}$~\Hunits~\cite{Corona:2021qxl}.

Concerning secret sterile neutrino interactions with a heavy mediator,
the situation for the $H_0$ tension is instead worsened
\cite{Forastieri:2017oma}.
When the sterile neutrino mass is enforced to obey SBL constraints and be close to 1~eV,
the study reports indeed that $H_0$ is even smaller in the self-interacting scenario than in the $\Lambda$CDM model,
because of the lower value of \Neff\ enforced by the interactions and the high value of $m_s$.
Since higher neutrino masses are correlated with lower $H_0$ values and a small \Neff\ also corresponds to a decrease in $H_0$,
both effects drive $H_0$ even down to $59.56 \pm 0.88$~\Hunits\ \cite{Forastieri:2017oma},
which of course increases the already strong $H_0$ tension.
This class of model is therefore ruled out by cosmological constraints on the sum of neutrino masses and on free-streaming of active neutrinos,
unless additional phenomena are considered to evade the neutrino mass bound, for example a fast decay mode from the sterile state into active neutrinos \cite{Chu:2018gxk}.

\section{Summary}
\label{sec:5}
Dark radiation models provide a natural environment where to explain many theoretical (QCD strong CP problem, origin of neutrino masses) and observational (short baseline neutrino oscillation anomalies) open problems in the Standard Model of elementary particles. Some of these new particles can also play a very important role in astrophysics and cosmology, e.g.\ contributing to the hot dark matter component of the universe, stellar evolution and more. 
The question we have explored throughout this review is whether the extra dark radiation component can also play a relevant role in resolving the so-called Hubble constant tension. 

The standard and commonly exploited way of parameterizing the mass-energy density in the new sector makes use of the \Neff\ parameter, the effective number of relativistic degrees of freedom. A larger value of \Neff\ can be obtained in very simple models, such as those with sterile neutrinos or with non-vanishing neutrino asymmetries. However, to solve the $H_0$ tension, more elaborated models leading to a non-zero $\DNeff$ are required. Interacting neutrino scenarios with a Majoron or with extra sterile species are examples in which the Hubble constant tension is alleviated. In addition, some of these models are also able to explain some of the aforementioned open questions within the SM, and therefore should be regarded as very appealing scenarios, both from the theoretical and observational perspectives. Upcoming data, not only from future cosmological probes, but also from laboratory, man-made particle beams and/or from astroparticle physics searches (neutrino, gamma-ray and cosmic ray telescopes in the case of axions and Majorons, for instance) may shed light on the role of extra dark radiation and the Hubble constant problem.

\begin{acknowledgement}
This work has been supported by the MCIN/AEI/10.13039/501100011033 of Spain under grant PID2020-113644GB-I00, by the Generalitat Valenciana of Spain under grant PROMETEO/2019/083 and by the European Union’s Framework Programme for Research and Innovation Horizon 2020 (2014–2020) under grant agreement 754496 (FELLINI) and 860881 (HIDDeN).
\end{acknowledgement}

%%%%%%%%%%%%%%%%%%%%%%%%%%%%%%%%%%%%%%%%%%%%%%%%%%%%%%%%%

% \bibliographystyle{JHEP}
\bibliography{biblio}

%%%%%%%%%%%%%%%%%%%%%%%%%%%%%%%%%%%%%%%%%%%%%%%%%%%%%%%%%

\end{document}